\documentclass[twocolumn,secnumarabic,amssymb,nobibnotes,aps]{revtex4-1}
\usepackage{graphicx}
\usepackage{dcolumn}
\usepackage{bm}
\usepackage{physics}
\usepackage{color}
\usepackage{float}
\usepackage{hyperref}

\renewcommand{\d}{\mathrm{d}}
\newcommand{\ee}{\mathrm{e}}
\newcommand*{\citen}{}
\DeclareRobustCommand*{\citen}[1]{%
  \begingroup
    \romannumeral-`\x 
    \setcitestyle{numbers}%
    \cite{#1}%
  \endgroup
}
\newcommand{\angles}[1]{%
	\left \langle #1 \right \rangle%
}

\begin{document}

\title{Correlation lengths in quasi-one-dimensional systems via transfer matrices}

\author{{Yi Hu\textit{$^{a}$}, Lin Fu\textit{$^{a}$} and Patrick Charbonneau$^{\ast}$\textit{$^{a,b}$}}\\
{\small \em $^a $Department of Chemistry, Duke University, Durham, North Carolina 27708, USA\\
$^b$Department of Physics, Duke University, Durham, North Carolina 27708, USA\\
$^*$ Email: patrick.charbonneau@duke.edu
}}

\date{\today}

\begin{abstract}
%

Using transfer matrices up to next-nearest-neighbour (NNN) interactions, we examine the structural correlations of quasi-one-dimensional systems of hard disks confined by two parallel lines and hard spheres confined in cylinders.
Simulations have shown that the non-monotonic and non-smooth growth of the correlation length in these systems accompanies structural crossovers (Fu \textit{et al., Soft Matter}, 2017, \textbf{13}, 3296).
Here, we identify the theoretical basis for these behaviour. In particular, we associate kinks in the growth of correlation lengths with eigenvalue crossing and splitting.
Understanding the origin of such structural crossovers answers questions raised by earlier studies, and thus bridges the gap between theory and simulations for these reference models.

\end{abstract}

\pacs{Valid PACS appear here}
\maketitle

\section{Introduction}

One-dimensional (1D) models capture the essence of experimental systems as varied as the trapping of cold atoms~\cite{kinoshita2004observation,moritz2005confinement,bloch2008many} and the self-assembly of metallic chains~\cite{shen1997magnetism,gambardella2002ferromagnetism}. They also are of keen theoretical interest because they lend themselves to exact analytical treatments~\cite{lieb2013mathematical}, and thus to insight into the nature of phase transitions in higher-dimensional systems. One way to partly bridge the gap between 1D and higher dimensions is to consider quasi-one-dimensional (q1D) models, which can be construed as strongly confined versions of higher-dimensional systems (with a single unbounded dimension)~\cite{yamchi2012fragile,ashwin2013inherent,godfrey2014static,godfrey2015understanding,yamchi2015inherent,robinson2016glasslike}. These models further have experimental analogues in the trapping of fullerenes within nanotubes~\cite{mickelson2003packing,khlobystov2004observation}, the self-assembly of nanoparticles within cylindrical pores~\cite{sanwaria2014helical,liang2014assembly,troche2005prediction} and the formation of colloidal wires~\cite{tymczenko2008colloidal}.

Just like their 1D counterparts, q1D models are amenable exact transfer-matrix treatments if confinement is sufficiently strong. (That is, as long as particles interact with but a small number of neighbours.) Transfer matrices were first used by Kramers and Wannier to solve the partition function of a 1D Ising chain~\cite{kramers1941statistics}, and 1D systems with up to third-nearest-neighbour (3NN) interactions have since been studied~\cite{hu2018clustering}. For q1D models, Barker first expressed the transfer-matrix formalism for systems in continuous space with both nearest-neighbour (NN) and next-nearest-neighbour (NNN) interactions~\cite{barker1962statistical}. Kofke and Post~\cite{kofke1993hard} first implemented this formalism for NN interactions to obtain the equations of state of hard disks confined between two parallel lines and of hard spheres in cylindrical pores~\cite{footnote1}.
Because the transfer matrix formalism involves solving an integral equation, however, the approach becomes rapidly intractable as the number of interacting neighbours increases. More specifically, the algorithmic complexity of the approach scales as $\mathcal{O}(N_{\rm g}^{2N_{\rm v}})$, where $N_{\rm g}$ is the number of grid points per dimension and $N_{\rm v}$ is the number of variables in the integral equation. Hence, only recently have systems with NNN interactions become computationally tractable~\cite{godfrey2015understanding,gurin2015beyond,gurin2017ordering}. Because Barker's formalism does not require that particles remain in the same order for all configurations, Varga \textit{et al.} have thus recently used it to study the behaviour of hard squares and rectangles confined between two parallel lines beyond the single-file condition~\cite{gurin2015beyond,gurin2017ordering}. Using a similar approach, Godfrey and Moore have studied similarly confined hard disks with NNN interactions~\cite{godfrey2015understanding}. The study of hard spheres with NNN interactions, however, has thus far been out of reach.

An important theoretical feature of q1D systems is that they do not present genuine phase transitions, as long as particle interactions remain short-ranged~\cite{van1950integrale,ruelle1999statistical,lieb2013mathematical}. As confinement weakens, these systems nonetheless exhibit a complex ordering behaviour characterized by relatively sharp structural crossovers~\cite{lohr2010helical,bogomolov1990sphere,jiang2013helical,koga2006close,duran2009ordering,huang2010direct,gordillo2006freezing,huang2009characterization,yue2011spontaneous,fu2016hard}.
A number of studies have shown that at finite pressure, these crossovers are accompanied with non-monotonic and non-smooth changes to the correlation length~\cite{godfrey2015understanding,fu2017assembly,gurin2017critical}.
Although the phenomenon is fairly general, its physical origin remains murky. In particular, it is unclear how can such behaviour of correlation length occur without accompanying phase transitions.

In this work, we calculate and compare different correlation lengths for both hard disks confined by two parallel lines and hard spheres confined in cylindrical pores using the NNN transfer-matrix formalism. In Section~\ref{sec:method} we describe the model and the formalism for the two systems, and in Section~\ref{sec:results} we present and discuss the relationship between correlation lengths and structural crossovers. We briefly conclude in Section~\ref{sec:conclude}.

\section{Transfer Matrix Method}
\label{sec:method}

Two types of q1D systems are considered in this work: hard disks of diameter $d$ confined between two parallel lines a distance $H$ apart, and hard spheres of diameter $d$ confined within a cylinder of diameter $D$. For notational convenience, we refer below to these two families of q1D systems as 2D and 3D systems, respectively.

\subsection{Systems with NN interactions} \label{sec:nn}
We first consider models with NN interactions alone. Choosing 2D systems with $1d<H < (1+\frac{\sqrt{3}}{2})d \approx 1.866 d$ enforces this condition. Without loss of generality, we choose the lines to be parallel to the $x$ axis and set the origin of the $y$ axis to lie halfway between those two lines. The available space for the disk centres in $y$ is then $h=H-d$, and the isobaric partition function for $N$ particles under the periodic boundary condition in the $x$ direction reads~\cite{kofke1993hard,varga2011structural}
\begin{eqnarray}
Z_{NPT} &&= \frac{1}{\Lambda^{2N}(\beta F)^{N}} \times \nonumber\\
&&\prod_{i=1}^{N}\int_{-h/2}^{h/2}\d y_i \exp\left[-\beta F\sum_{j=1}^{N}\sigma_2(y_j,y_{j+1})\right],
\end{eqnarray}
where $\Lambda$ is the thermal de Broglie wavelength, $F=P_xh$ is a force associated with the axial pressure $P_x$, and $\sigma_2(y,y')=\sqrt{d^2-(y-y')^2}$ is the contact distance between two neighbouring particles along the $x$ direction.

Defining the kernel
\begin{equation}
K_{2,\mathrm{NN}}(y,y')\equiv{\rm exp}\left[-\beta F\sigma_2(y,y')\right],
\end{equation}
allows us to rewrite the configurational part of the partition function as $\Tr(K_{2,\mathrm{NN}}^N)$, where the trace of $K_{2,\mathrm{NN}}$ is the sum over all eigenvalues $\lambda_k$ raised to the $N$-th power that satisfy
\begin{equation} \label{eq:NN2deig}
\int_{-h/2}^{h/2} \d y K_{2,\mathrm{NN}}(y,y')u_k(y)=\lambda_k u_k(y'),
\end{equation}
where $u_k(y)$ is the eigenfunction corresponding to $\lambda_k$.
In the thermodynamic $N\rightarrow\infty$ limit, $\Tr(K_{2,\mathrm{NN}}^N)$ is dominated by the largest eigenvalue, $\lambda_0^N$. In this case, the Gibbs free energy per particle is
\begin{equation}
\beta G/N = \lim_{N\rightarrow\infty}\frac{-\ln Z_{NPT}}{N}= 2\ln\Lambda + \ln(\beta F)-\ln\lambda_0,
\end{equation}
the equation of state for the linear density of particles, $\rho_{\rm l}=N/L$, is
\begin{equation}
\begin{aligned}
\label{eq:density}
\frac{1}{\rho_{\rm l}} &= \frac{1}{\beta F}- \left(\frac{\partial {\rm ln}\lambda_0}{\partial\beta F}\right)_{\beta,N} = \frac{1}{\beta F}+\frac{1}{\lambda_0}\times \\
&\int_{-h/2}^{h/2}\!\!\!\d y u_0(y)\int_{-h/2}^{h/2}\!\!\!\d y'u_0(y') K_{2,\mathrm{NN}}(y,y')\sigma_2(y,y')
\end{aligned}
\end{equation}
and the probability distribution function of the $y$ coordinate is $u^2_0(y)$. 

The spatial correlation between the $y$ coordinate of two particles is then
\begin{eqnarray}
g_y(i,j) &=& \langle(y_i-\langle y_i\rangle)(y_j-\langle y_j\rangle)\rangle\nonumber\\
&=&\langle y_iy_j\rangle-\langle y_i\rangle\langle y_j\rangle=\langle y_iy_j\rangle,
\label{eq:gy}
\end{eqnarray}
with $\langle y_i\rangle =0$ by symmetry. In terms of the kernel eigenvalues and eigenfunctions, we have for a translationally invariance system~\cite{varga2011structural}
\begin{eqnarray}
g_y(i, j)&=&g_y(|i-j|)\nonumber\\
&=&\sum_{k=1}^{\infty}\left(\frac{\lambda_k}{\lambda_0}\right)^{|i- j|}\left[\int_{-h/2}^{h/2}\d yu_0(y)yu_k(y)\right]^2\nonumber\\
&=&\sum_{k=1}^{\infty}\left(\frac{\lambda_k}{\lambda_0}\right)^{|i- j|}I_k^2,
\label{eq:xi}
\end{eqnarray}
where the integral $I_k$ was implicitly defined. Because such correlations are expected to decay exponentially at large distances, i.e., $g_y(n)\sim {\rm exp}(-n/\xi_y)$ with correlation length $\xi_y/\rho_{\rm l}$ for $n\rightarrow\infty$, we have $\xi_y^{-1}=\mathrm{ln}(\lambda_0/|\lambda_1|)$, where $\lambda_{1}$ is the second largest eigenvalue (if $I_1\neq0$).  

For 3D systems, the same formalism allows us to obtain the partition function in terms of cylindrical coordinates--radial $\sqrt{\varrho}$ and angular $\varphi$ components--~\cite{kofke1993hard}
\begin{equation}
\begin{aligned}
&Z_{NPT} = \frac{1}{\Lambda^{3N}(\beta F)^{N}} \prod_{i=1}^{N} \int_{0}^{w^2}\!\!\!\frac{\d \varrho_i}{2} \times \\
&\int_{-\pi}^{\pi}\!\!\!\d\varphi_i \exp\left[-\beta F\sum_{j=1}^{N}\sigma_3(\varrho_j,\varrho_{j+1},\varphi_j,\varphi_{j+1})\right],
\end{aligned}
\end{equation}
where $w=(D-d)/2$, $F=P_x\pi (D/2)^2$ and 
\begin{equation}
\sigma_3(\varrho,\varrho',\varphi,\varphi')=\sqrt{d^2-\varrho-\varrho'+2\sqrt{\varrho\varrho'}{\rm cos}(\varphi-\varphi')}.
\end{equation}
Because the contact separation $\sigma_3$ depends on the relative angular separation $\psi = \varphi - \varphi'$. The kernel can be simplified as
\begin{equation}
K_{3,\mathrm{NN}}(\varrho,\varrho',\psi)\equiv{\rm exp}\left[-\beta F\sigma_3(\varrho,\varrho',\psi)\right],
\end{equation}
with corresponding eigenequation 
\begin{eqnarray}
(\hat{H} u_k)(\varrho',\psi')&=&\int_{0}^{w^2} \d \varrho \int_{-\pi}^\pi \d \psi K_{3,\mathrm{NN}}(\varrho,\varrho',\psi) u_k(\varrho,\psi)\nonumber\\
&=&\lambda_k u_k(\varrho',\psi').
\label{eq:h}
\end{eqnarray}
The angular component can be integrated separately~\cite{kofke1993hard}, hence Eq.~\eqref{eq:h} can be simplified further as
\begin{equation}
\int_{0}^{w^2} \d \varrho K'_3(\varrho,\varrho') u_{0,r}(\varrho)=\lambda_0 u_{0,r}(\varrho')\label{eq:k3}
\end{equation}
with
\begin{equation}
K'_{3,\mathrm{NN}}(\varrho,\varrho')=\int_{-\pi}^{\pi} K_{3,\mathrm{NN}}(\varrho,\varrho',\psi) \d \psi.
\end{equation}
This simplification thus expresses the equation of state of this model from the solution of a single one-dimensional eigenproblem.

The spatial correlation for the radial component of the 3D system system can then be extracted from
\begin{eqnarray}
&&g_r(i,j) = \langle r_i r_j\rangle-\langle r_i\rangle\langle r_j\rangle=\langle r_i r_j\rangle-\langle r_j\rangle^2\\
&&=g_r(|i-j|) = \sum_{k=1}^{\infty}\left(\frac{\lambda_k}{\lambda_0}\right)^{|i- j|}\left[\int_{0}^{w^2} u_{0,r}(\varrho) \sqrt{\varrho} u_k(\varrho)\right]^2,\nonumber
\label{eq:gtheta}
\end{eqnarray}
and the solutions, $\lambda_k$ and $u_k(\varrho)$, give $\xi_r^{-1}=\ln(\lambda_0/|\lambda_1|)$. Note that because the eigenproblem in Eq.~\eqref{eq:k3} is but a simplification of Eq.~\eqref{eq:h}, the eigenvalues of $K_{3,\mathrm{NN}}$ and $K'_{3,\mathrm{NN}}$ are the same. The correlation length $\xi_r$ can therefore be obtained from diagonalizing $K_{3,\mathrm{NN}}$ directly. 

\subsection{Systems with NNN interactions} \label{sec:nnn}

\begin{figure}[h]
\centering
  \includegraphics[width=0.9\columnwidth]{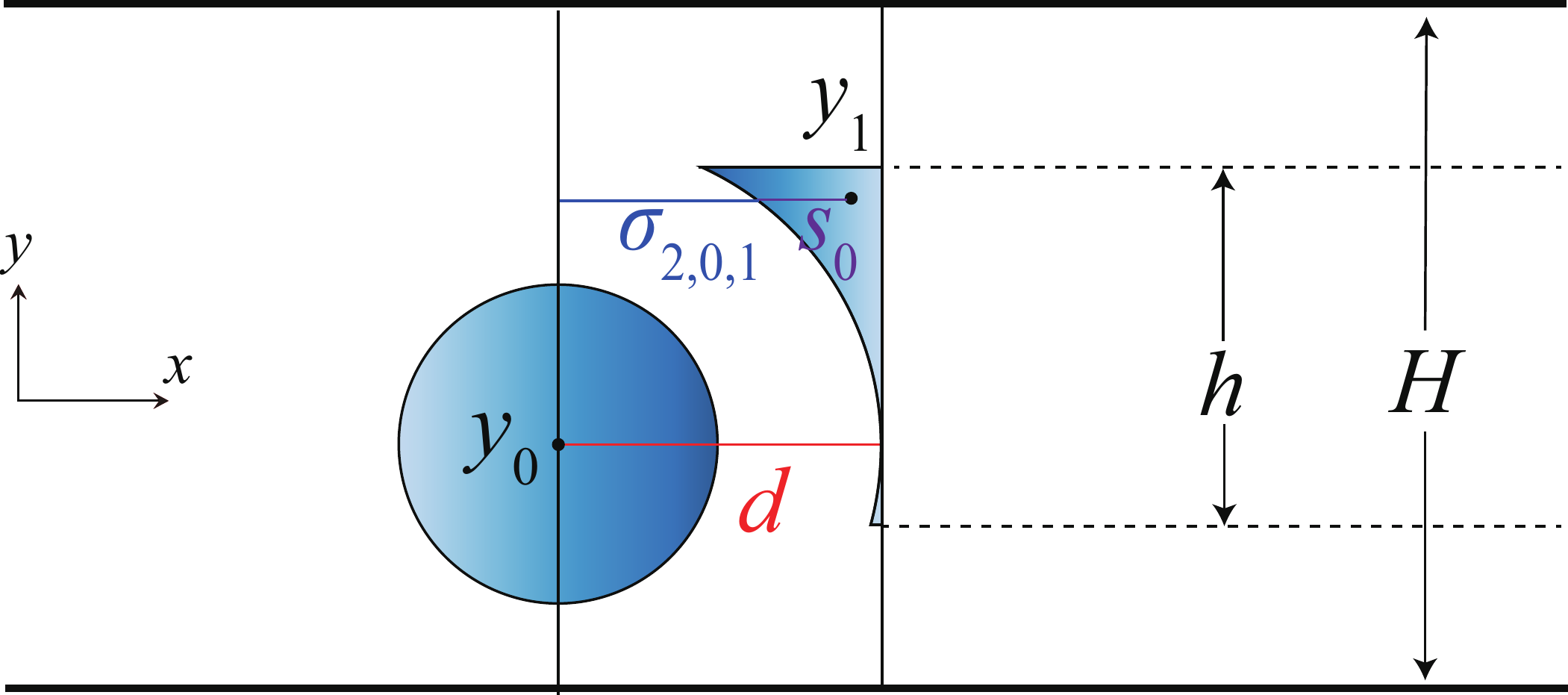}
  \caption{Illustration of the NNN integration variables for a 2D system. The line separation, $H$, and the available height $h=H-d$, for hard disks of diameter $d$. The $y$ coordinates of two nearby disks, $y_0$ and $y_1$, set the contact separation along the $x$ axis of $\sigma_{2,0,1}$. The excess separation, $s_0$, is the axial separation from contact. Disk $2$ (not shown) may contact disk $0$ only if the centre of disk $1$ is within the shaded area.}
    \label{fig:2ddemonstrate}
\end{figure}

For 2D systems in the NNN regime, i.e. for $(1+\frac{\sqrt{3}}{2}) d \le H < 2.5 d$, the formalism of Ref.~\citen{godfrey2015understanding} gives
\begin{align}
\label{eq:nnn}
&Z_{NPT}=\frac{1}{\Lambda^{2N}} \prod_{i=1}^N \times \\
&\int_0^\infty\!\!\! \d s_i \ee^{-\beta Fs_i}\int_{-h/2}^{h/2}\!\!\!\d y_i\ee^{-\beta F\sum_{j=1}^{N-1}\sigma_2(y_j,y_{j+1})}\prod_{k=1}^{N-2}\Theta_k,\nonumber
\end{align}
where the separation between neighbouring disks along the $x$ axis is denoted
\begin{equation}
\begin{aligned}
s_i &= x_{i+1}-x_{i}-\sigma_2({y_i,y_{i+1}}) \\
&= \Delta x_i - \sigma_2({y_i,y_{i+1}}),
\end{aligned}
\end{equation}
and the NNN interactions are expanded in terms of Heaviside step functions
\begin{eqnarray}
\Theta_k&\equiv&\theta(s_k+s_{k+1}+\sigma_{2,k,k+1}+\sigma_{2,k+1,k+2}-\sigma_{2,k,k+2})\nonumber\\
&=&\theta(x_{k+2}-x_k-\sigma_{2,k,k+2}).
\end{eqnarray}
Here, we write $\sigma_{2,i,j}\equiv\sigma_2(y_i,y_j)$ for short. From this point on, the shorthand subscripts on quantities are used to denote relative particle indices, e.g. $y_{i}, y_{i+1}$, and $y_{i+2}$ are simply $y_0, y_1$, and $y_2$, respectively.

Following the same procedure as for NN interactions, we define the NNN kernel,
\begin{equation} \label{eq:NNN2Dkernel}
K_{2,\mathrm{NNN}}(y_0,y_1,y_2,s_0,s_1)={\rm exp}[-\beta F \Delta x_0]\Theta_0,
\end{equation}
and solving the partition function is equivalent to solving the eigenequation
\begin{eqnarray} \label{eq:NNN2deig}
&&\int_{-h/2}^{h/2}\d y_2 \int_0^\infty \d s_1  K_{2,\mathrm{NNN}} u'_k(y_1,y_2,s_1)\nonumber\\
&&=\lambda_k u'_k(y_0,y_1,s_0),
\label{eq:2dnnn}
\end{eqnarray}
which gives the free energy $\beta G/N=2\ln \Lambda-\ln \lambda_0$ in the thermodynamic limit.

If we consider the NN interaction regime using the NNN formalism for a moment, we note that the step functions in Eq.~\eqref{eq:NNN2Dkernel} are then all unity~\cite{godfrey2015understanding}. This simplification is what enables integrating the axial part separately. $K_{2,\mathrm{NNN}}$ thus reduces back to $K_{2,\mathrm{NN}}$ in this regime. Because the eigenvalues of $K_{2,\mathrm{NN}}$ and $K_{2,\mathrm{NNN}}$ are the same, so are their associated correlation lengths. Therefore, in addition to $\xi_y^{-1}={\rm ln}(\lambda_{0}/|\lambda_1|)$, we can generally investigate the correlation length for the axial separation between neighbouring disks, $\xi_{\Delta x}^{-1}={\rm ln}(\lambda_0/|\lambda_2|)$, in both NN and NNN regimes, as suggested in Ref.~\citen{godfrey2015understanding}.

The equation of state derivation is similar to Eq.~\eqref{eq:density}, specifically,
\begin{align} \label{eq:NNN2Ddensity}
\frac{1}{\rho_{\rm l}}&=\int_{-h/2}^{h/2}\!\!\!\d y_0\int_{-h/2}^{h/2}\!\!\!\d y_1 \int_0^\infty\!\!\! \d s_0 \Delta x_0 \times \nonumber \\
&\ee^{-\beta F\Delta x_0} u'_0(y_0,y_1,s_0)u'_0(y_1,y_0,s_0)/B[u'_0,u'_0]
\end{align}
with
\begin{eqnarray}
B[u'_k,u'_m]&=&\int_{-h/2}^{h/2}\!\!\!\d y_0\int_{-h/2}^{h/2}\!\!\!\d y_1\int_0^\infty\!\!\! \d s_0 \ee^{-\beta F \Delta x_0} \times \nonumber \\
&&u'_k(y_0,y_1,s_0)u'_m(y_1,y_0,s_0).
\end{eqnarray}

The treatment of the NNN regime in 3D, i.e. $(1+\frac{\sqrt{3}}{2}) d \le H < 2 d$~\cite{mughal2012dense}, is akin to that of 2D. Replacing the eigenfunctions $u'_k(y_0,y_1,s_0)$ by $u'_k(\varrho_0,\varrho_1,\psi_0,s_0)$, the kernel reads
\begin{equation}
\begin{aligned}
\label{eq:nnn3dkernel}
  K_{3,\mathrm{NNN}}(\varrho_0,\varrho_1,\varrho_2,\psi_0,\psi_1,s_0,s_1) =  \exp[-\beta F \Delta x_0] \Theta_0.
\end{aligned}
\end{equation}

\subsection{Relationship between simple structural correlation and axial distribution function}

It is important, however, to note that the correlation lengths defined in Sections~\ref{sec:nn} and \ref{sec:nnn} characterize the asymptotic behaviour of the correlation functions between pairs (or pairs of pairs) of particles with index differences, $|i-j|$, as independent variables. In standard liquid state theory, by contrast, correlation lengths typically refer to the spatial decay of correlations based on interparticle distances. To relate the latter with the former, one can write down an expansion in terms of distribution of $n$-th nearest neighbours,
$
g(r) = \sum_{n=1}^{N_\mathrm{ubound}+1} g_n(r),
$
where $N_\mathrm{ubound}$ is the maximum number of particles that fit between a pair of particles with axial separation $r$. The closed form of $g_n(r)$ is then
\begin{equation} \label{eq:spatialcor}
\begin{aligned}
g_n(r) &= \frac{1}{\rho_{\rm l} Z_{nPT}} \int Z_{nPT} \delta(r' - r) \prod_{i=0}^{n-1} d \bm{x}_i \\
&= \frac{1}{\rho_{\rm l} Z_{nPT}} \int \delta(r' - r) \prod_{i=0}^{n-1} (K_\mathrm{NNN}(\bm{x}_i) \d \bm{x}_{i}) \\
&= \frac{1}{\rho_{\rm l} Z_{nPT}} \ee^{-\beta F r} \int \delta(r' - r) \prod_{i=0}^{n-1} \Theta_i \d \bm{x}_i
\end{aligned}
\end{equation}
where $\bm{x}_{i}$ denotes the coordinates of a single particle, $\delta$ is the Dirac delta function, and $Z_{nPT}$ normalizes $\rho_{\rm l} \int_0^\infty g_n(r) \d r = 1$. The integration over space, has $r'$ as the axial separation between the first ($0$-th) and the last ($n$-th) particles, such that $r' = \sum_{i=0}^{n-1} \Delta x_i$. 

To the best of our knowledge, a closed form expression that would generally relate $g_n(x)$ with $g_{\Delta x}(n)$ is not known. In the limit $r \rightarrow \infty$, however, $g_n(r)$ can be approximated as a normal distribution near its maximum, $g_n(r) \sim \mathcal{N}(n\rho_{\rm l}, \mathrm{Var}(g_n(r)))$, with variance
\begin{equation}
\begin{aligned}
\mathrm{Var}(g_n(r)) &= \angles{ (\sum_{i=0}^{n-1} (\Delta x_i - \angles{\Delta x}))^2 } \\
&= n \mathrm{Var}(\Delta x) + \sum_{i=1}^{n-1} 2 (n-i) g_{\Delta x}(i), \\
\end{aligned}
\end{equation}
where $\mathrm{Var}(\Delta x) = \angles{(\Delta x - \angles{\Delta x})^2} = g_{\Delta x}(0)$.
This approximation reveals that the axial distribution function is intimately related to the variance of $\Delta x$ and to simple correlation functions. The non-monotonic changes to the correlation lengths of the axial distribution function reported in Ref.~\citen{fu2017assembly} should thus be echoes of similar phenomena in $g_{\Delta x}$. 


\subsection{Numerical solution} \label{sec:num}

The integral equations in Section~\ref{sec:nnn} can be solved numerically after discretizing the arguments of the kernel. Here, we adapt the approach of Ref.~\citen{kofke1993hard} for the NN regime, hence only the approach for the NNN regime is detailed below. 

For the 2D NNN case, discretizing Eq.~\eqref{eq:2dnnn} gives
\begin{eqnarray}
\delta y\delta s\sum_{i=1}^{n_y}\sum_{j=1}^{n_s}K_{2,\mathrm{NNN}}(y_0,y_1,i\delta y,s_0,j\delta s) \nonumber\\
\times u'_k(y_1,i\delta y,j\delta s)=\lambda_k u'_k(y_0,y_1,s_0).
\label{eq:dis}
\end{eqnarray}
Solving a series of Eq.~\eqref{eq:dis} for all $(y_0, y_1, s_0)$ is therefore equivalent to solving the matrix eigenvalue problem 
$\mathbf{K} \bm{u} = \lambda \bm{u}$. Although this operation is not possible for arbitrarily high limits of integration on $s$, the Tonks-like exponential decay of $u'_k(y,y',s)$ beyond $s_\mathrm{max}(y_0, y_1)=d-\sigma_2(y_0,y_1)$ gives $u'_k(y_0,y_1,s_0)=u'_k(y,y',s_\mathrm{max})\ee^{-\beta F(s_0-s_\mathrm{max})}$ for $s_0>s_\mathrm{max}$. The integration beyond $s_\mathrm{max}$ can thus be performed analytically. We here choose the discretization interval of $s$ by assigning the maximal possible discretization number of the axial separation $n_{s,\mathrm{max}}$, such that
\begin{equation}
\begin{aligned}
 \delta s &= 1 / (n_{s,\mathrm{max}}-1) \max_{\{y_0, y_1\}} s_\mathrm{max}(y_0, y_1) \\
 &= s_\mathrm{max}(-\frac{h}{2}, \frac{h}{2}) / (n_{s,\mathrm{max}}-1).
\end{aligned}
\end{equation}
The number of discretized $s$ values for a given pair of $y_0$ and $y_1$ is then
\begin{equation}
 n_s(y_0, y_1) = \lceil s_\mathrm{max}(y_0, y_1)  / \delta s \rceil + 1.
\end{equation}
Figure~\ref{fig:2ddemonstrate} illustrates the range of $s$ values needed for this construction as the shaded area near the hard disk.

At high pressures, disks are found close to the confining walls with high probability, and thus both the probability density and $u'_k(y,y')$ have most of their weight near these walls. The effect is especially pronounced for $h>\frac{\sqrt{3}}{2}d$. 
In order to minimize the number of grid points along the $y$ coordinates ($n_y$) needed to obtain a given accuracy, we perform the change of variables suggested by Ref.~\citen{godfrey2015understanding},
\begin{equation}
 y(t) = at+b{\rm tanh}(ct),
\end{equation}
which gives $y(\pm1)=\pm h/2$ and $\frac{\d y}{\d t}=a+b\ c\ \mathrm{sech}^2(ct)$, where $b=h/2$, $a=h/2-b\mathrm{tanh}(c)$.
Using a larger $c$ refines the grid close to the confining walls.
Here, we find that $c=3$ for $h<\sqrt{3}/2$ and $c=6$ for $h \ge \sqrt{3}/2$ provide sufficient numerical accuracy. The range $t \in [-1, 1]$  is then uniformly discretized with $n_t = n_y$.

In summary, the entries of the transfer matrix are
\begin{widetext}
\begin{equation} \label{eq:NNN2Dmat}
\begin{aligned}
\left[\mathbf{K}\right]_{(y_0, y_1, s_0),(y_1', y_2, s_1)} = 
\begin{cases}
0, & y_1 \neq y_1' \\
\delta t \delta s \frac{\d y}{\d t} \ee^{-\beta F (\sigma_{2,0,1}+s_0)} \Theta_0, & y_1 = y_1' \text{ and } s_0 < s_\mathrm{max}(y_0, y_1) \\
\delta t \frac{\d y}{\d t} \ee^{-\beta F (\sigma_{2,0,1}+s_0)}/(\beta F), & y_1 = y_1' \text{ and } s_0 = s_\mathrm{max}(y_0, y_1)
\end{cases}
\end{aligned}
\end{equation}
\end{widetext}
where the subscripts denote the row and column indices. The elements of a given row $(y_0, y_1, s_0)$ are thus either the same nonzero value, $K(y_0, y_1, s_0)$, or $0$. Unless otherwise specified, we choose $n_y=100$ for $H \le (1+\frac{\sqrt{3}}{2})d$ and $n_y=150$ for $H > (1+\frac{\sqrt{3}}{2})d$ with $n_{s,\mathrm{max}}=100$.

The construction of Eq.~\eqref{eq:NNN2Dmat} generates a structured and highly-singular matrix.
Its form can thus be simplified in order to reduce the algorithmic complexity.
The element-wise form of matrix-vector multiplication $\bm{w} = \mathbf{K} \bm{v}$ reads
\begin{equation} \label{eq:NNN2Drow}
\begin{aligned}
w(y_0, y_1, s_0) &= K(y_0, y_1, s_0) \sum_{\{y_2\}} \sum_{\{s_1(y_1, y_2)\}} v(y_1, y_2, s_1) \Theta_0.
\end{aligned}
\end{equation}
The complexity of obtaining one element is $\mathcal{O}(n_y n_{s,\mathrm{max}})$, and the operation must be done for $\mathcal{O}(n_y^2 n_{s,\mathrm{max}})$ elements, hence the overall complexity is $\mathcal{O}(n_t^3 n_{s,\mathrm{max}}^2)$. Calculating  and storing the list of
\begin{equation}
  \tilde{v}(y_1, y_2, s_1) = \sum_{s \in [s_1, s_{\mathrm{max}}(y_1, y_2)]} v(y_1, y_2, s)
\end{equation}
with $\mathcal{O}(n_y^2 n_{s,\mathrm{max}})$ operations reduces the complexity of obtaining one element of Eq.~\eqref{eq:NNN2Drow} to $\mathcal{O}(n_y)$. Furthermore, if NNN interactions are not possible for a given combination of $(y_0, y_1, s_0)$ with $(y_1,y_2, s_1)$ in which $y_2$ and $s_1$ are arbitrary, then calculating $w(y_0, y_1, s_0')$ for $s_0' > s_0$ only requires $\mathcal{O}(1)$ operations, because $w(y_0, y_1, s_0')$ is then $w(y_0, y_1, s_0)$. For $H < (1+\frac{\sqrt{3}}{2}) d$, these optimizations reduce the computational complexity to $\mathcal{O}(n_y^3 + n_y^2 n_{s,\mathrm{max}})$; for $H \ge (1+\frac{\sqrt{3}}{2}) d$, the complexity approaches $\mathcal{O}(n_y^3 n_{s,\mathrm{max}})$ in the large $n_{s,\mathrm{max}}$ limit.

Similarly, the entries of the transfer matrix for 3D systems in the NNN regime are
\begin{widetext}
\begin{equation} \label{eq:NNN3Dmat}
\begin{aligned}
\left[\mathbf{K}\right]_{(\varrho_0, \varrho_1, \psi_0, s_0),(\varrho_1', \varrho_2, \psi_1, s_1)} = 
\begin{cases}
0, & \varrho_1 \neq \varrho_1' \\
\delta \varrho \delta \psi \delta s \ee^{-\beta F (\sigma_{3,0,1}+s_0)} \Theta_0, & \varrho_1 = \varrho_1' \text{ and } s_0 < s_\mathrm{max}(\varrho_0, \varrho_1, \psi_0) \\
\delta \varrho \delta \psi \ee^{-\beta F (\sigma_{3,0,1}+s_0)}/(\beta F), & \varrho_1 = \varrho_1' \text{ and } s_0 = s_\mathrm{max}(\varrho_0, \varrho_1, \psi_0) \\
\end{cases}
\end{aligned}
\end{equation}
\end{widetext}
and the calculation of $\bm{w} = \mathbf{K} \bm{v}$ can be written term by term as
\begin{align} \label{eq:NNN3Drow}
w&(\varrho_0, \varrho_1, \psi_0, s_0) = \\
&K(\varrho_0, \varrho_1, \psi_0, s_0) \sum_{\{\varrho_2\}} \sum_{\{\psi_1\}}\sum_{\{s_1(\varrho_1, \varrho_2)\}} v(\varrho_1, \varrho_2, s_1) \Theta_0. \nonumber
\end{align}
By applying the same optimizations as for 2D NNN systems, the complexity of matrix-vector multiplication of the kernel can be reduced from $\mathcal{O}(n_\varrho^3 n_\psi^2 n_s^2)$ to $\mathcal{O}(n_\varrho^3 n_{\psi}^2 + n_\varrho^2 n_{\psi} n_{s,\mathrm{max}})$ for $D<(1+\frac{\sqrt{3}}{2}) d$, and $\mathcal{O}(n_\varrho^3 n_{\psi}^2 n_{s,\mathrm{max}})$ for $D>(1+\frac{\sqrt{3}}{2}) d$ in the large $n_{s,\mathrm{max}}$ limit. We here choose $n_\varrho=n_{\psi}=n_{s,\mathrm{max}}=100$. A typical calculation of the first three eigenvalues then takes less than a day in Matlab~\cite{Matlab2016a}.

For both 2D and 3D systems, the densities given by Eqs.~\eqref{eq:density} and \eqref{eq:NNN2Ddensity} can be rewritten in matrix form
\begin{equation} \label{eq:numdensity}
\begin{aligned}
\frac{1}{\rho_{\rm l}} &=  -\left(\pdv{\ln \lambda_{0}}{\beta F}\right)_{\beta} 
  = - \frac{\bm{u}^{-1} (\partial{\mathbf{K}}/\partial(\beta F))_\beta \bm{u}}{\bm{u}^{-1} \bm{u} \lambda_{0}}
\end{aligned}
\end{equation}
where $\bm{u}$ and $\bm{u}^{-1}$ are the right and left eigenvectors corresponding to $\lambda_0$, respectively.
Because $\mathbf{K}$ is not symmetric, we have that $(\bm{u}^{-1})^T \neq \bm{u}$, and thus the row vector $\bm{u}^{-1}$ shall be obtained by solving another eigenproblem, $\mathbf{K}^T (\bm{u}^{-1})^T = \lambda (\bm{u}^{-1})^T$.
Similar optimization as above can be made in computing $\bm{w}'=\mathbf{K}^T \bm{v}'$. In 2D system with NNN interactions, for instance, one can calculate and store a list of 
\begin{equation}
 \tilde{v}'(y_0, y_1, s_0) = \sum_{s \in [s_0, s_{\mathrm{max}}(y_0, y_1)]} K(y_0, y_1, s) v'(y_0, y_1, s),
\end{equation}
and make use of the fact that $w'(y_1, y_2, s_1')=w'(y_1, y_2, s_1)$ if NNN interactions are not possible for any $s_1'>s_1$. In the end, this formalism has the same complexity as the calculation of $\mathbf{K} \bm{v}$.

Correlation functions can also be obtained by transfer matrices. For instance, the matrix form of Eq.~\eqref{eq:xi} reads
\begin{equation}
 g_y(|i-j|) = \angles{y_i y_j} = \frac{(\bm{y}^T \circ \bm{u}^{-1}) \mathbf{K}^{|i-j|} (\bm{y} \circ \bm{u})}{\bm{u}^{-1} \bm{u} \lambda_{0}^{|i-j|}},
\end{equation}
where $\bm{y}$ is a column vector with $y(y_0, y_1, s_0) = y_0$ and ``$\circ$'' denotes the Hadamard product. Once the correlation function has been computed, $\xi_y$ can also be obtained by fitting $\ln g_y(|i, j|) = -|i-j|/\xi_y + \mathrm{constant}$. 

\begin{figure}[h]
\centering
  \includegraphics[width=0.9\columnwidth]{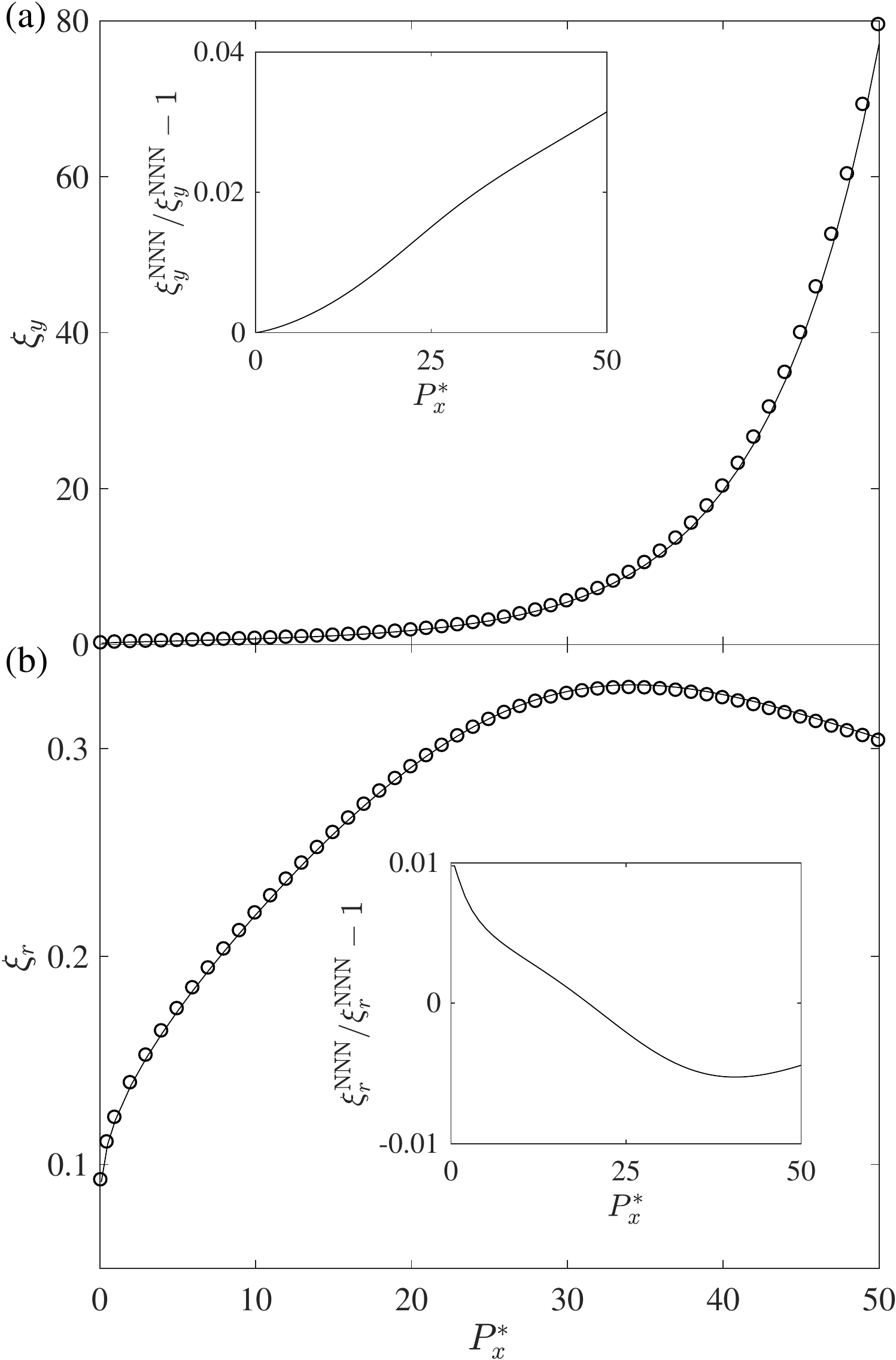}
  \caption{Comparison between the NN (lines) and NNN (circles) correlation lengths for (a) 2D systems with $H=1.5d$, and (b) 3D systems with $D=1.5d$. (Insets) The relative deviations between the NNN and NN formalisms are less than 3\% over the range of pressures considered.}
    \label{fig:nn_nnn}
\end{figure}

The above numerical approach was first validated by successfully comparing the resulting equation of state with those of previous simulations~\cite{gordillo2006freezing,fu2017assembly} (not shown).
We also compared the NN and the NNN calculations of $\xi_y$ in 2D and $\xi_r$ in 3D systems with each other and with previous results~\cite{varga2011structural}. Agreement between the two schemes is excellent at low pressures, and discrepancies of at most $\sim3\%$ develop as pressure increases (Fig.~\ref{fig:nn_nnn}a). This mismatch results from the growing sharpness in the form of the eigenfunctions at high pressure, which reduces the numerical accuracy of the NNN calculation. In both cases, however, the error is quantitatively fairly modest and should not qualitatively alter our analysis.



\section{Results and Discussion}
\label{sec:results}

\begin{figure*}[t]
  \centering
  \includegraphics[width=0.95\textwidth]{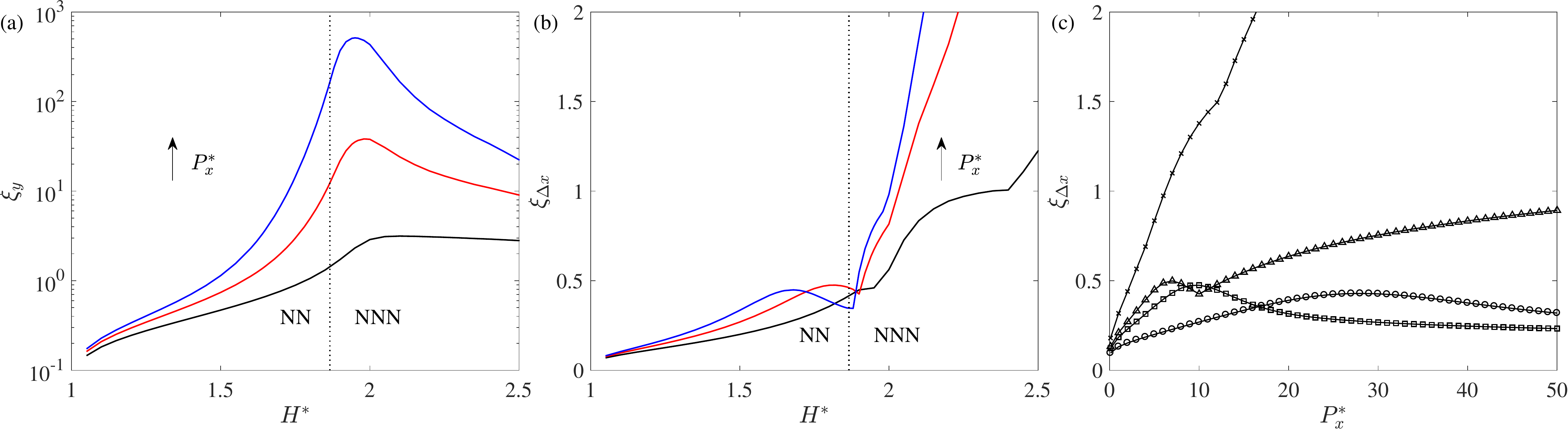}
  \caption{(a) Evolution of $\xi_y$ with $H^*=H/d$ for $P_x^*=5$ (black), $10$ (red) and $15$ (blue); (b) Evolution of $\xi_{\Delta x}$ with $H^*$, following the same notation as in (a); (c) Evolution of $\xi_{\Delta x}$ with $P_x^*$ for $H^*=1.5$ (circles), $1.8$ (squares), $1.9$ (triangle) and $2.1$ (crosses).}
    \label{fig:2d}
\end{figure*}

In this section, the numerical results for various correlation lengths in 2D and 3D systems are reported and discussed within the broader context of ordering in q1D models. 

\subsection{2D Systems}
In 2D systems, we investigate the behaviour of two correlation lengths, $\xi_y$ and $\xi_{\Delta x}$ (Fig.~\ref{fig:2d}). $\xi_{y}$ is known to grow monotonically with (unitless) pressure, $P_x^*=\beta F d$, and to diverge nearly exponentially as $P_x^* \rightarrow \infty$~\cite{varga2011structural,godfrey2015understanding}. Here, we find that the behaviour of $\xi_y$ is nonetheless non-trivial in some ways. Under a fixed intermediate pressure, $\xi_y$ indeed shows a peak near $H^*=H/d=2$. Because $\xi_y$ quantifies the spatial decay of correlations between particles, its decrease for $H^*>2$ corresponds to an increase in the probability of configurational defects.
$\xi_{\Delta x}$ presents an even richer behaviour, evolving non-monotonically with both $P_x^*$ and $H$. 
A previous study found that the non-monotonic behaviour of $\xi_{\Delta x}$ at high pressures corresponds to the structural crossover between zig-zag and buckled zig-zag order. However, the low-pressure regime, in which the straight-chain to zig-zag crossover should occur, has not previously been considered.

\begin{figure}[h!]
\centering
  \includegraphics[width=0.9\columnwidth]{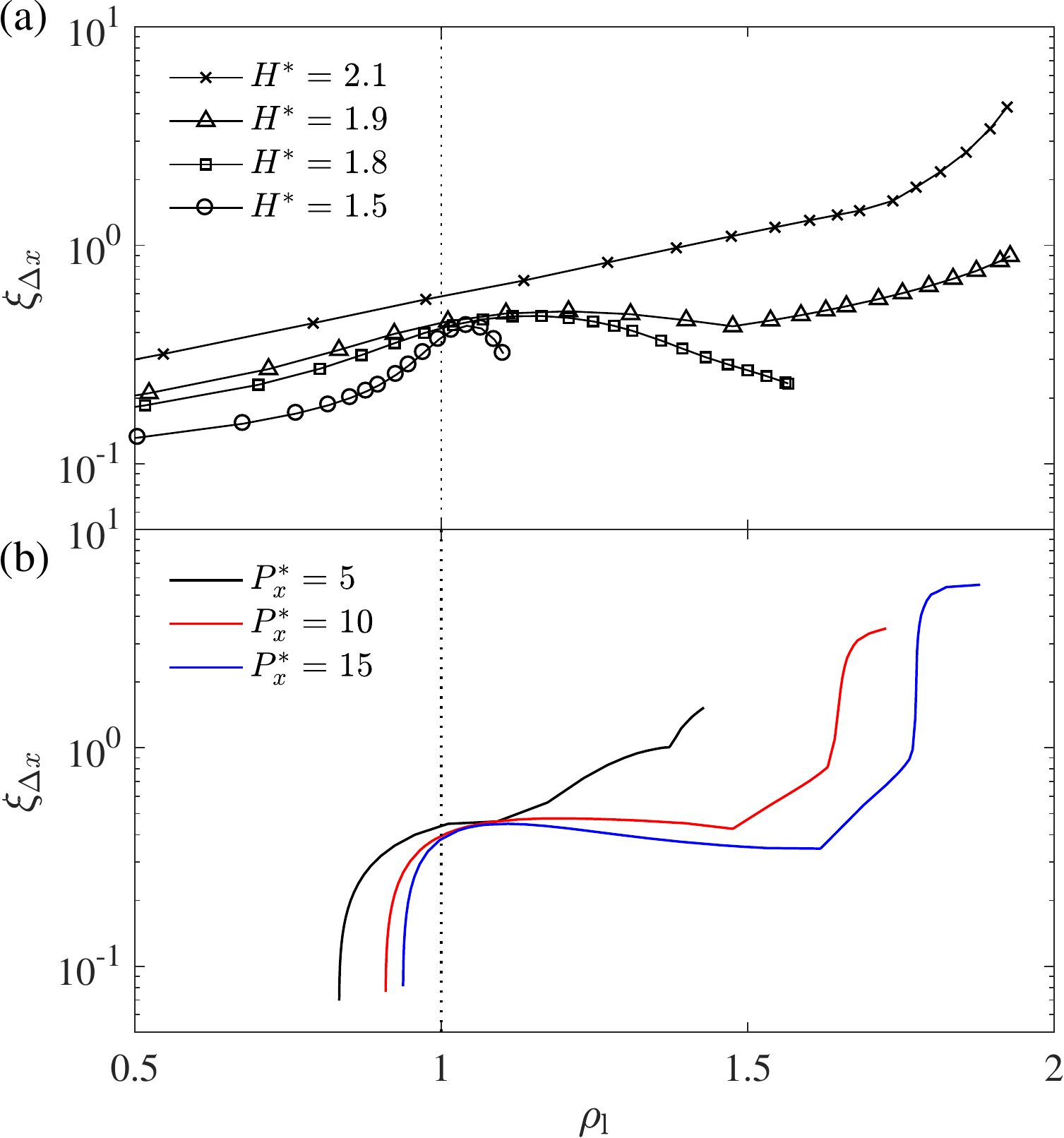}
  \caption{Parametric plot of the evolution of $\xi_{\Delta x}$ with (a) $P_x^*$ and (b) $H^*$. For a fixed $H^*$ (or $P_x^*$), increasing $P_x^*$ (or $H^*$) increases $\rho_{\rm l}$. The correspondence between $H^*$ and $P_x^*$ is obvious; all curves for $H^*<2$ peak around $\rho_{\rm l}=1$, at which point the system transforms from straight-chain to zig-zag order.}
    \label{fig:para}
\end{figure}

To confirm the existence of a structural crossover, we consider the evolution of $\xi_{\Delta x}$ with $\rho_{\rm l}$. We find that $\xi_{\Delta x}$ peaks around $\rho_{\rm l}\approx1$ (Fig.~\ref{fig:para}), which is the maximum linear density of a straight chain. Systems with $\rho_{\rm l}>1$, must therefore display significant zig-zag order. For $\rho_{\rm l}<1$, particles can freely fluctuate between the two lines, and order grows smoothly with density increasing.
For $\rho_{\rm l}>1$, however, zig-zag order grows but axial correlations decrease.
To better understand this point consider, for instance, a perfect zig-zag structure, which is the densest structure for $H < (1+\frac{\sqrt{3}}{2}) d$. It has $ \Delta x_i - \angles{\Delta x}=\sigma_2-\sigma_2=0$, $\forall i$, hence $g_{\Delta x}(n)\equiv0$, $\forall n$ and $\xi_{\Delta x}=0$. In other words, the correlation function vanishes because fluctuations are then suppressed.
Hence, although the system is strongly correlated at high pressures, $\xi_{\Delta x}$ does not measure the range of correlations in the NN regime, although it does sensitively capture the presence of the structural crossover associated with the onset of zig-zag order. This explains why the smaller the line separation the more significant this crossover, and why it is absent when zig-zag order is no longer stable, i.e.,  $H^* \ge 2$. Note that a peak in $\xi_{\Delta x}$ around $\rho_{\rm l}=1$ also appears upon varying the wall separation at fixed $P_x^*$. Choosing $\rho_{\rm l}$ as independent parameter, the evolution of $\xi_{\Delta x}$ with both $P_x^*$ and $H^*$ thus displays the same non-monotonic behavior.

\begin{figure*}[t!]
\centering
  \includegraphics[width=0.95\textwidth]{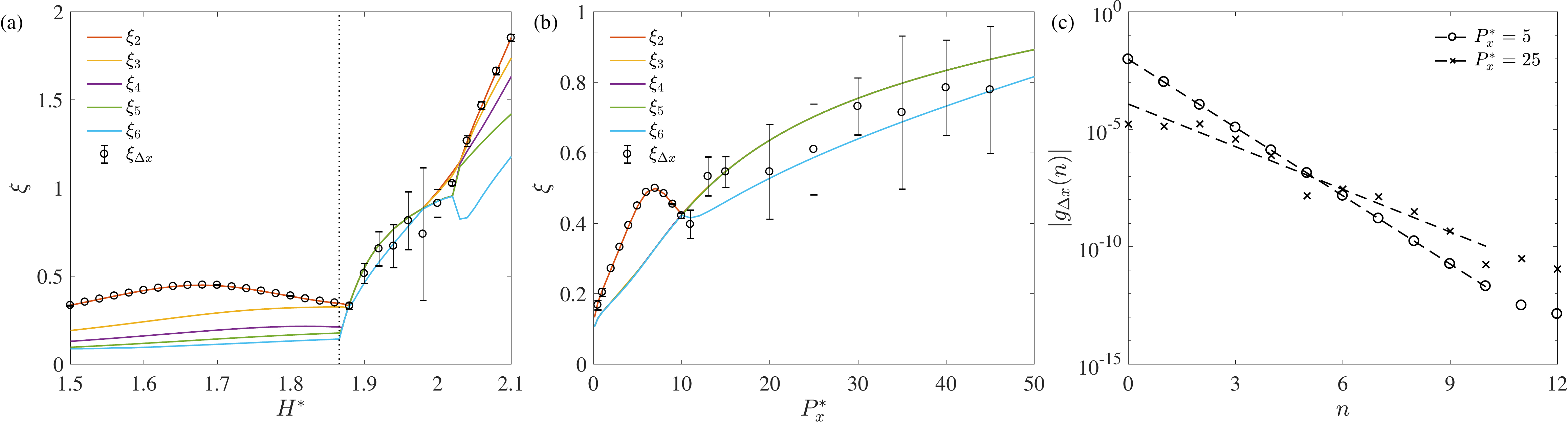}
  \caption{(a, b) The $i$-th largest correlation lengths obtained from $\xi_i = 1/\ln(\lambda_0/|\lambda_i|)$ compared with $\xi_{\Delta x}$ estimated from fitting $g_{\Delta x}(n)$ (circles) for (a) varying $H^*$, with $P_x^*=15$ and (b) varying $P_x^*$, with $H^*=1.9$ in 2D systems. Error bars denote the $95\%$ confidence intervals. (c) Exponential decay of $g_{\Delta x}(n)$ at $H^*=1.9$ for $P_x^*=5$ (circles) and $25$ (crosses) in lin-log scale. The latter clearly shows complex oscillation behaviour which increases the uncertainty of estimate of $\xi_{\Delta x}$.}
    \label{fig:2DHxi}
\end{figure*}

Near the onset of the NNN regime, kinks can be observed in the growth of $\xi_{\Delta x}$. At this onset, new types of structural correlations appear and grow quickly, and hence once one of these new correlations starts dominating (in this case, the fourth eigenvalue becomes the third), $\xi_{\Delta x}$ forms a kink (Fig.~\ref{fig:2DHxi}a). Physically, this signals that the main order type changes from zig-zag to zig-zag with (NNN specific) defects, as described in Ref.~\cite{ashwin2009complete}. A kink can be observed in the pressure dependence of $\xi_{\Delta x}$ for the same exact reason (Fig.~\ref{fig:2DHxi}b). 

Interestingly, the numerical uncertainty of obtaining $\xi_{\Delta x}$ by fitting the decay of $g_{\Delta x}$ directly also grows after the kink (Fig.~\ref{fig:2DHxi}c). When $1.88 \le H^* \le 1.98$ under $P_x^*=15$ or $P_x^*>10$ under $H^*=1.9$, for instance, $\pm\lambda_2$ and their complex conjugates (four equal correlation lengths, $\xi_2 \sim \xi_5$) all give $\xi_{\Delta x}$. The early decay of the correlation function is then more complex than a simple exponential, which explains the relatively large size of the error bars.

For $H^* \gtrsim2$, another eigenvalue crossing appears. This one coincides with the crossover in $\xi_y$ discussed above. As the zig-zag configuration becomes no longer stable and new configurations emerge at $H^*>2$, this kink is here again due to a change of dominant correlation type. Note that beyond this kind $|\lambda_2|$ is no longer degenerate, hence oscillations in $g_{\Delta x}$ are once again minimal and so is the error on fitting the associated correlation length. 



\subsection{3D Systems}

\begin{figure*}[t!]
\centering
  \includegraphics[width=0.95\textwidth]{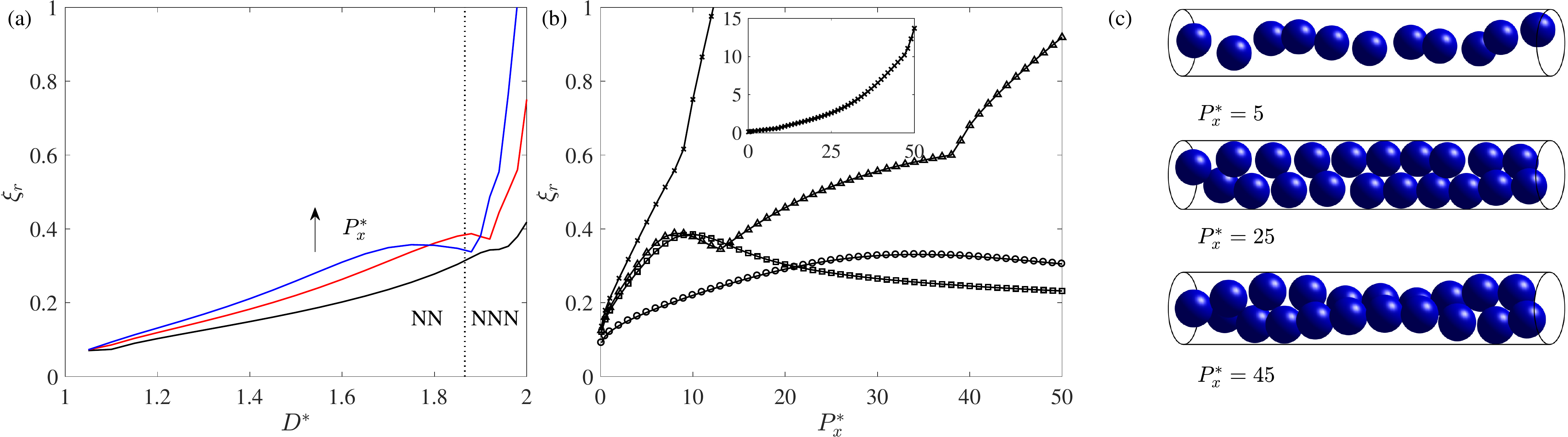}
  \caption{(a) Evolution of $\xi_r$ with $D$ at $P_x^*=5$ (black), $10$ (red), and $15$ (blue). (b) Evolution of $\xi_r$ with $P_x^*$ at $D^*=D/d=1.5$ (circles), $1.866$ (squares), $1.9$ (triangle) and $2$ (cross). (inset) Extended $\xi_r$ range for $D^*=2$. (c) Typical configurations for hard spheres in cylindrical pore with $D^*=1.9$ obtained in simulations~\cite{fu2017assembly}. The system is gas-like at $P_x^*=5$, displays zig-zag order at $P_x^*=25$, and exhibits significant helicity at $P_x^*=45$.}
    \label{fig:3d}
\end{figure*}

\begin{figure}[h!]
\centering
  \includegraphics[width=0.72\columnwidth]{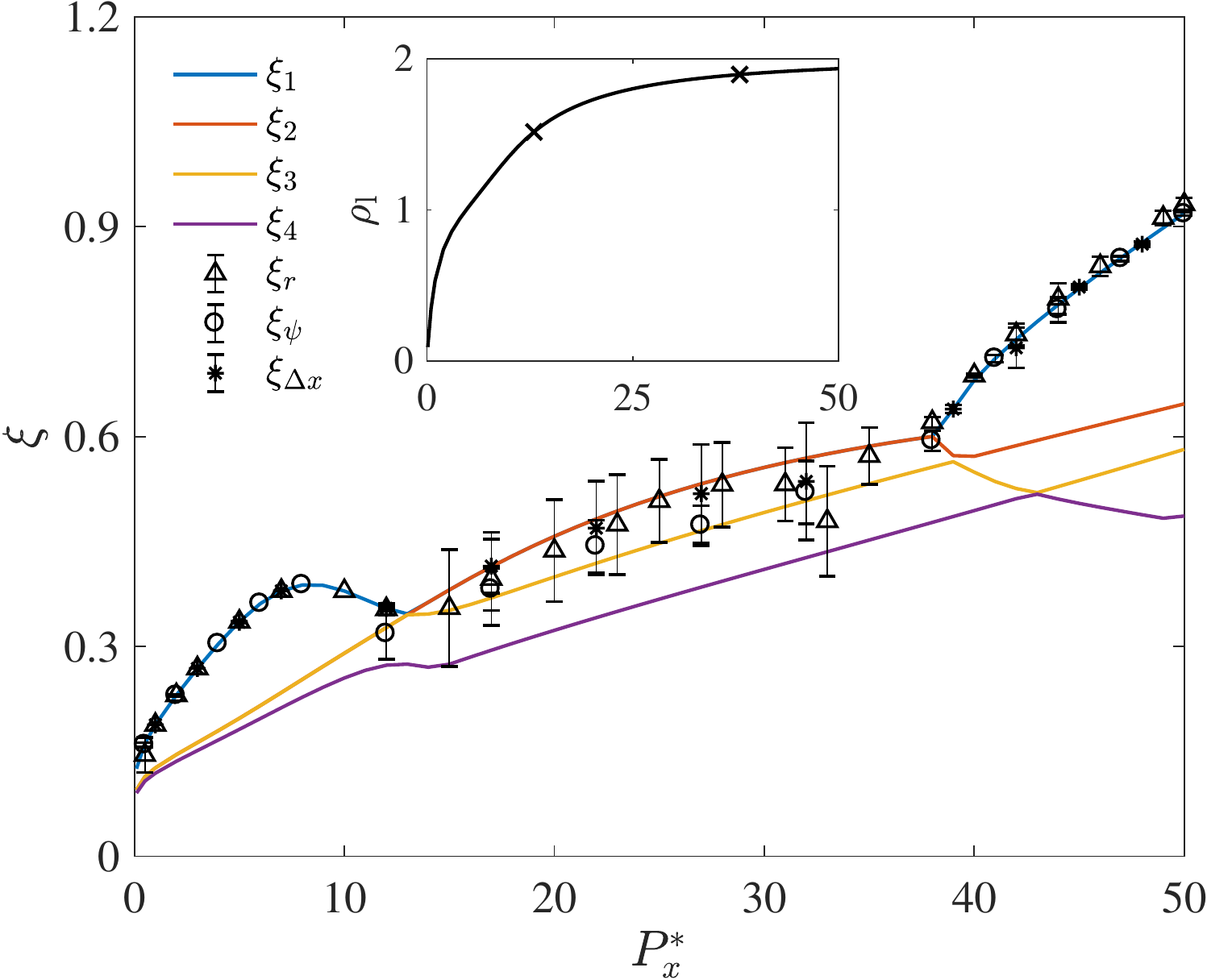}
  \caption{The $i$-th largest correlation lengths obtained from $\xi_i = 1/\ln(\lambda_0/|\lambda_i|)$ compared with $\xi_r$, $\xi_{\psi}$, $\xi_{\Delta x}$ estimated from fitting $g(n)$ for varying $P_x^*$, with $D^*=1.9$ for 3D systems. The error bars denote the range of $95\%$ confidence intervals for the estimates as in Fig.~\ref{fig:2DHxi}. (inset) Equation of state of the same system. The densities at which kinks appear in $\xi_r$ are marked as crosses, but correspond to no detectable feature in the equation of state.}
    \label{fig:3DPxi}
\end{figure}

As shown in Fig.~\ref{fig:3d}a, the growth of the correlation length with cylinder diameter $D$ is similar to that $\xi_{\Delta x}$ in 2D. The non-monotonicity of $\xi_r$ is here as well associated with the structural crossover between straight chain and zig-zag order. The kinks that appear around $D=(1+\frac{\sqrt{3}}{2})d$ also indicates the onset of NNN contributions to the correlation. Beyond this point, $\xi_r$ grows monotonically until $D=2d$, at which point interactions beyond NNN become possible. This similarity is physically intuitive for $D<(1+\frac{\sqrt{3}}{2})d$, because helicity is then weak in 3D and the densest packings are essentially the same as in 2D~\cite{pickett2000spontaneous}. Correspondingly, in this regime $\xi_r$ first grows with pressure and then decreases (Fig.~\ref{fig:3d}b), just like $\xi_{\Delta x}$ in 2D. Figure~\ref{fig:3DPxi}a also shows that a notable uncertainty in fitting correlation lengths is observed in this regime, but seems to be relatively less than that in 2D as the degeneracy is here double rather than quadruple. 

In 3D, however, an extra kink appears at high pressures. In this case, the two conjugate eigenvalues at intermediate pressure become two different real eigenvalues. Based on the observations in Fig.~\ref{fig:3d}c, this kink appears to be related to the crossover from zig-zag to compact helical order. The absence of a comparable kink in 2D systems supports this hypothesis.
Note that this change does not leave any signature in the equation of state (Fig.~\ref{fig:3DPxi}a, inset).
The behavior of the correlation length is thus clearly a more sensitive measure of the structural crossovers.

As an additional comment, while it is mathematically demonstrated that in 2D that $\xi_y$ and $\xi_{\Delta x}$ are not equivalent~\cite{godfrey2015understanding}, we could obtain no such demonstration for 3D. In fact, as also observed in Ref.~\citen{fu2017assembly} we find here that the correlation lengths $\xi_\psi$ and $\xi_{\Delta x}$ behave equivalently to $\xi_r$. The underlying reason thus remains nebulous.

\section{Conclusion} \label{sec:conclude}

We have examined the behaviours of different correlation lengths in systems of hard disks confined between two parallel lines and hard spheres confined within a cylindrical pore with up to NNN interactions.
The correlation length, $\xi_{\Delta x}$, which probes axial order in 2D, grows non-monotonically in both the NN and the NNN regimes. This non-monotonicity accompanies the crossover from straight chain to zig-zag order. 
$\xi_{\Delta x}$ also displays kinks, which are due to eigenvalue crossing as NNN interactions arise.
These phenomena are also observed in 3D, but hard spheres systems additionally show a kink associated with the onset of helicity.
In general, both 2D and 3D systems have richer ordering behaviour as confinement weakens, which explains the complex results obtained in simulation of q1D systems.


\section*{Acknowledgments} 
We thank Sho Yaida for stimulating discussions.
We also thank Daan Frenkel for his unwavering support throughout the years. We wish him a very happy 70th birthday.
This work was supported by the National Science Foundation's grant from the Nanomanufacturing Program (CMMI-1363483).

\nocite{*}
\bibliography{abbrev}

\end{document}